\documentclass[twocolumn, superscriptaddress, prb, nofootinbib]{revtex4}
\usepackage{graphicx}
\usepackage{amsmath}
\usepackage{multirow}
\usepackage{textcomp}
\usepackage{float}
\usepackage{hyperref}
\usepackage[symbol]{footmisc}
\usepackage{perpage}

\begin{document}
\title{Impact of Intermediate Sites on Bulk Diffusion Barriers: Mg Intercalation in Mg$_2$Mo$_3$O$_8$}

\author{Gopalakrishnan Sai Gautam} \email{gautam91@mit.edu}
\thanks{G. S. Gautam and X. Sun contributed equally to the work}
\affiliation{
Department of Materials Science and Engineering, Massachusetts
Institute of Technology, Cambridge, MA 02139, USA}
\affiliation{
Materials Science Division, Lawrence Berkeley National Laboratory,
Berkeley, CA 94720, USA}
\affiliation{
Department of Materials Science and Engineering, University of California Berkeley,
CA 94720, USA}

\author{Xiaoqi Sun}
\thanks{G. S. Gautam and X. Sun contributed equally to the work}
\affiliation{
Department of Chemistry and the Waterloo Institute of Nanotechnology, University of Waterloo, ON N2L3G1, Canada}

\author{Victor Duffort}\affiliation{
Department of Chemistry and the Waterloo Institute of Nanotechnology, University of Waterloo, ON N2L3G1, Canada}

\author{Linda F. Nazar}\affiliation{
Department of Chemistry and the Waterloo Institute of Nanotechnology, University of Waterloo, ON N2L3G1, Canada}

\author{Gerbrand Ceder} \email{gceder@berkeley.edu, gceder@lbl.gov}
\affiliation{
Materials Science Division, Lawrence Berkeley National Laboratory,
Berkeley, CA 94720, USA}
\affiliation{
Department of Materials Science and Engineering, University of California Berkeley,
CA 94720, USA}



\begin{abstract}
The ongoing search for high voltage positive electrode materials for Mg batteries has been primarily hampered by poor Mg mobility in bulk oxide frameworks. Motivated by the presence of Mo$_3$ clusters that can facilitate charge redistribution and the presence of Mg in a non-preferred (tetrahedral) coordination environment, we have investigated the Mg (de)intercalation behavior in layered-Mg$_2$Mo$_3$O$_8$, a potential positive electrode. While no electrochemical activity is observed, chemical demagnesiation of Mg$_2$Mo$_3$O$_8$ is successful but leads to amorphization. Subsequent first-principles calculations predict a strong thermodynamic driving force for structure decomposition at low Mg concentrations and high activation barriers for bulk Mg diffusion, in agreement with experimental observations. Further analysis of the Mg diffusion pathway reveals an O--Mg--O dumbbell intermediate site that creates a high Mg$^{2+}$ migration barrier, indicating the influence of transition states on setting the magnitude of migration barriers.
\end{abstract}

\maketitle
Rechargeable Mg batteries have received interest as an energy storage system that potentially offers high energy density. The major advantage relies on the benefits of Mg metal as the negative electrode, which, in addition to being inexpensive, abundant and safe in handling and storage, also provides high volumetric capacity (3833~mAh~cm$^{-3}$) and can be free of dendrite growth when operating in an electrochemical cell.\cite{Muldoon2014, Yoo2013, Muldoon2012}  However, the development of corresponding positive electrode materials has been slow.\cite{Yoo2013} Since the discovery of the first seminal functional Mg insertion positive electrode -- the Chevrel phase (CP, Mo$_6$S$_8$),\cite{Aurbach2000} only recently have two other structures been shown to be suitable for Mg (de)intercalation in a full cell arrangement with a Mg anode, namely the spinel and layered titanium sulfide.\cite{Sun2016,Sun2016a} The above materials take the advantage of a ``soft" anionic framework that interacts weakly with the Mg$^{2+}$ and assist its mobility. On the contrary, sluggish multivalent ion mobility is generally observed in oxide lattices. Nevertheless, oxides are still of great interest due to their potentially higher operating voltage.\cite{Amatucci2001,Levi2009,Rong2015,SaiGautam2015,Bo2015,Gershinsky2013,Gautam2015,SaiGautam2016,Liu2015}

\begin{figure}[h]
\centering
  \includegraphics[width=\columnwidth]{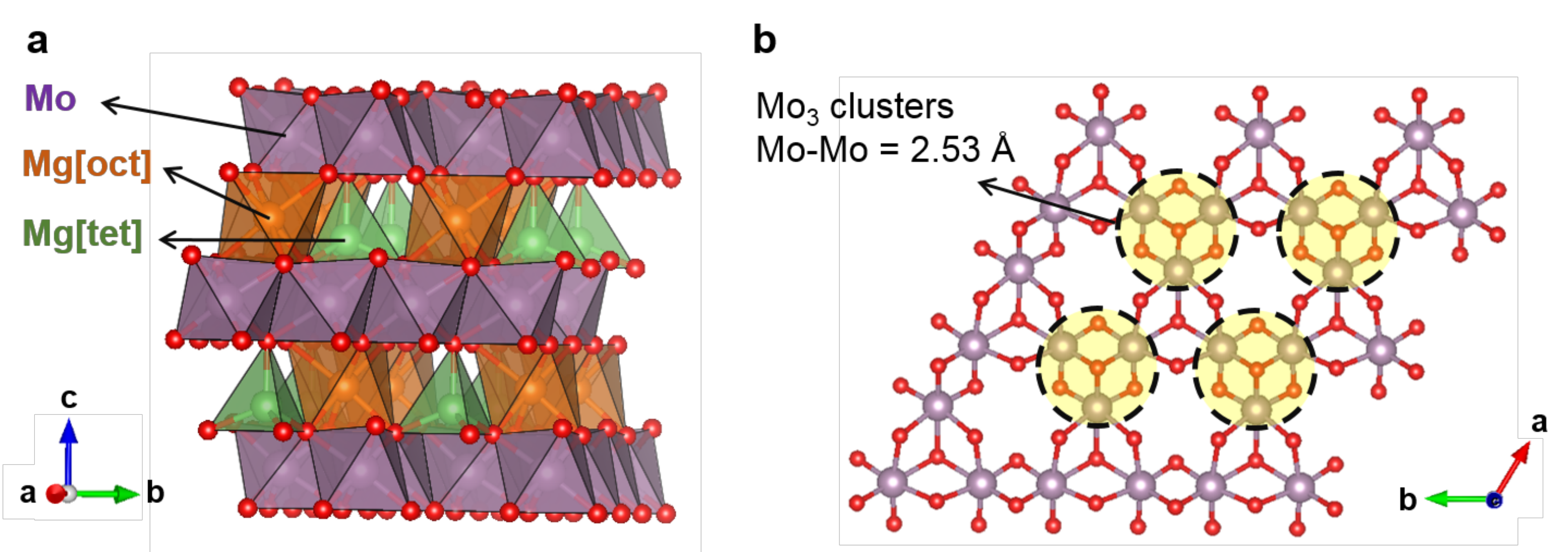}
  \caption{(a) Crystal structure of Mg$_2$Mo$_3$O$_8$. (b) MoO$_6$ octahedra layer showing Mo$_3$ clusters.}
  \label{fig:1}
\end{figure}

Levi \emph{et al.} have speculated that the presence of Mo$_6$ clusters in the CP structure is one of the key factors for facile Mg$^{2+}$ mobility by promoting charge redistribution.\cite{Levi2009}  The possibility that a similar principle may apply to oxides guided us to Mg$_2$Mo$_3$O$_8$ (Figure~\ref{fig:1}a), which has Mo$_3$ clusters in the Mo$_3$O$_8$ layers (Figure~\ref{fig:1}b).\cite{McCarroll1957,Cotton1964} In this structure, Mg occupies both octahedral and tetrahedral sites between the layers.  While octahedral Mg share both edges and corners with MoO$_6$ octahedra (Figure~\ref{fig:1}a), tetrahedral Mg share corners with MoO$_6$ and MgO$_6$ octahedra. Since Mg is also present in a ``non-preferred" tetrahedral coordination (Figure~\ref{fig:1}a),\cite{Brown1988}  a Mg diffusion pathway with lower migration barriers is expected than when Mg is exclusively found in its preferred octahedral coordination,\cite{Rong2015} such as in conventional layered oxides. In the case of layered oxides, the Mg diffusion pathways contain an intermediate tetrahedral site presumably with high energy relative to the stable octahedral site, leading to poor Mg mobility.\cite{Rong2015}

We note that the Li analogue (Li$_4$Mo$_3$O$_8$) has previously been examined in a Li cell, offering 218~mAh~g$^{-1}$ initial specific capacity.\cite{Barker2005}  Other materials with similar structures containing Mo$_3$ clusters, such as LiMoO$_2$ and Li$_2$MoO$_3$, also function well as Li-ion positive electrodes.\cite{Barker2005,Ben-Kamel2012,Ma2014}  On the other hand, only limited work has been done for Mg intercalation in Mo-oxides,\cite{Gershinsky2013,Incorvati2015,Kaveevivitchai2016} motivating us to examine the Mg$^{2+}$ diffusion properties in Mg$_2$Mo$_3$O$_8$ and its potential to be a positive electrode material for Mg batteries.

\begin{figure}[h]
\centering
  \includegraphics[width=\columnwidth]{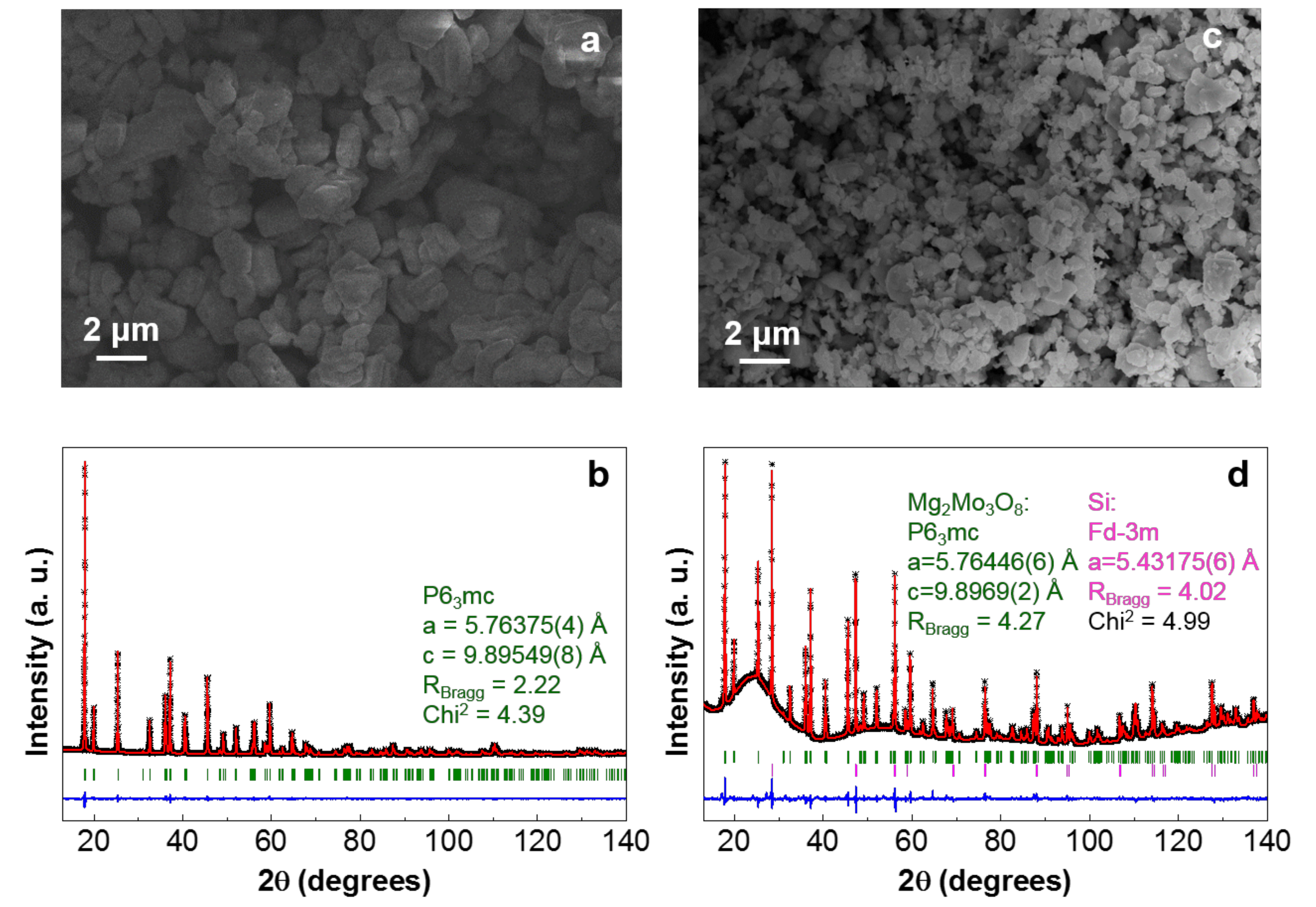}
  \caption{(a) SEM image and (b) Rietveld refinement fit of pristine Mg$_2$Mo$_3$O$_8$ (Bragg-Brentano geometry). (c) SEM image of the demagnesiated sample. (d) Rietveld refinement of the demagnesiated sample with external silicon standard added to evaluate the percentage of amorphous phase. The mixture was sealed in X-ray capillary under Ar and was measured in Debye-Scherrer geometry. Black crosses~--~experimental data, red lines~--~fitted data, blue line~--~difference map between observed and calculated data, green ticks~--~the \textit{P6$_3$mc} phase of Mg$_2$Mo$_3$O$_8$, pink ticks~--~the \textit{Fd$\bar{3}$m} phase of Si.}
  \label{fig:2}
\end{figure}

Mg$_2$Mo$_3$O$_8$ was obtained by solid-state synthesis (see Electronic Supporting Information -- ESI\footnote{$^\#$ Electronic Supporting Information available free of charge online at \url{http://dx.doi.org/10.1039/C6TA07804D}}$^{\#}$) and provided particles a few micrometers in size (Figure~\ref{fig:2}a). Its X-ray diffraction (XRD) pattern was indexed in the \textit{P6$_{3}$mc} space group characteristic of this material (Figure~\ref{fig:2}b). In order to study the possibility of Mg removal from such a structure, chemical demagnesiation was carried out using NO$_2$BF$_4$, a commonly used oxidizing agent for chemical delithiation.\cite{Wizansky1989}   Mg$_2$Mo$_3$O$_8$ and NO$_2$BF$_4$ were reacted in a 1:4 ratio, which would allow complete Mg de-intercalation if each NO$_2$BF$_4$ sustained a one electron reduction as anticipated. Energy dispersive X-ray spectroscopy (EDX) reveals that the majority of the Mg was removed from the structure (Table~\ref{tab:1}). The particles become smaller after demagnesiation (Figure~\ref{fig:2}c), suggesting some changes in the material. Despite these differences, the XRD results indicate no shift of the peaks (Figure~\ref{fig:2}d). The atomic positions obtained by Rietveld refinement\cite{Rietveld1969}  are almost the same as the pristine (Table~S1 in ESI), suggesting that a two-phase reaction takes place, with the demagnesiated phase being amorphous. During this process, Mg is presumably first removed from the outer shell, leading to the destabilization of the parent lattice and eventual amorphization. The amount of amorphous phase in the demagnesiated product is estimated to be around 87 wt\% using Si as an external standard method (Figure~\ref{fig:2}d, see ESI for details), giving an overall composition of Mg$_{0.24}$Mo$_3$O$_8$, which is similar to the cationic ratio determined by EDX (Table~\ref{tab:1}) and indicates complete demagnesiation of the amorphous component. The $\sim$~13 wt\% unreacted Mg$_2$Mo$_3$O$_8$ results from the reduced oxidizing strength of NO$_2$BF$_4$ exhibited near the end of the reaction due to low oxidizer concentration, or other side reactions. Partial demagnesiation from Mg$_2$Mo$_3$O$_8$ was not achieved when the ratio of oxidizing agent was reduced (Mg$_2$Mo$_3$O$_8$:NO$_2$BF$_4$~=~1:2), as indicated by the preservation of the initial phase obtained by XRD refinement (Figure~S1a and Table~S1c). Together with the decrease of overall Mg concentration (Mg/Mo ratio of $\sim$~0.53(4)/3 by EDX) and the co-existence of different morphologies (Figure~S1b), the XRD data suggests that part of Mg$_2$Mo$_3$O$_8$ undergoes complete demagnesiation and becomes amorphous with some fraction of material not participating in the reaction.

\begin{table}[h]
\small
  \caption{\ EDX results of Mg$_2$Mo$_3$O$_8$ before and after chemical demagnesiation.}
  \label{tab:1}
  \begin{tabular*}{0.48\textwidth}{@{\extracolsep{\fill}}lll}
    \hline
    Sample & Pristine & Demagnesiated \\
    \hline
    Mg/Mo & 1.59(4)/3 & 0.13(6)/3 \\
    \hline
  \end{tabular*}
\end{table}

Since the degree of chemical oxidation was hard to control, we attempted to evaluate stepwise demagnesiation behavior by an electrochemical method. As it has been suggested that the Mg desolvation process depends on the solvent,\cite{Canepa2015,Canepa2015a}  and this is critical for the electrochemical mechanism at the positive electrode,\cite{Wan2015,Sun2016b}  Mg$_2$Mo$_3$O$_8$ was examined in both non-aqueous (all phenyl complex~--~APC\cite{Mizrahi2008}) and aqueous (Mg(ClO$_4$)$_2$ in water) systems. A demagnesiation voltage similar to the delithiation of Li$_4$Mo$_3$O$_8$ (average of $\sim$~2.4~V vs.\ Mg),\cite{Barker2005} or at $\sim$~2.6~V as predicted by first principles calculations (Figure~S4, see ESI for details) could be expected. Both electrolytes offer a stable voltage window for this range; however, no electrochemical activity was observed in either system (Figure~S2). Such results potentially indicate the existence of a high Mg$^{2+}$ diffusion barrier in the structure, hence kinetics being the main limitation. Chemical oxidation, on the other hand, might involve a mechanism other than simple cation diffusion, such as a partial dissolution/re-precipitation process. This helps in lowering the kinetic barrier and establishes successful Mg removal.

In order to understand the amorphization upon chemical demagnesiation and rationalize the lack of electrochemical activity in Mg$_2$Mo$_3$O$_8$, we carried out first principles calculations to determine the energy above hull (E$^{\rm hull}$) indicating the stability of the structure, and the activation barriers for Mg diffusion within the structure (methodological details of the calculations are provided in the ESI).


The energy above the convex ground state hull (E$^{\rm hull}$) of the Mg$_{\rm x}$Mo$_3$O$_8$ structure, calculated with respect to the stable compounds in the Mg-Mo-O ternary phase diagram, can be used to evaluate the thermodynamic stability of the structure on demagnesiation.\cite{Gautam2015,Liu2015} Typically, a thermodynamically stable structure will have an E$^{\rm hull}$ of 0~meV/atom, while more positive E$^{\rm hull}$ values indicate greater driving force to form other phases, which may be reflected as difficulty in synthesizing a compound, or as decomposition during (de)intercalation. Also, E$^{\rm hull}$ values are evaluated at 0~K and entropic contributions can stabilize a structure at higher temperatures. The values listed in Table~\ref{tab:2} have been determined from the available compounds in the Materials Project database.\cite{Jain2013}  The trends in Table~\ref{tab:2} indicate an increasing E$^{\rm hull}$ with increasing Mg removal from the Mg$_2$Mo$_3$O$_8$ structure, corresponding to an increase in the thermodynamic driving force for decomposition. The E$^{\rm hull}$ values at lower Mg concentrations are very high -- consistent with the experimentally observed amorphization during chemical Mg extraction from Mg$_2$Mo$_3$O$_8$ (Figure~\ref{fig:2}) and the naturally amorphous occurrence of Mo$_3$O$_8$.\cite{Vlek1977,Staples1951} 

\begin{table*}[b]
\small
  \caption{\ The E$^{\rm hull}$ values (in meV/atom) and the corresponding decomposition products are listed as a function of Mg content in the Mo$_3$O$_8$ structure, as obtained from the Materials Project database. The comments column indicates available experimental observations.}
  \label{tab:2}
  \begin{tabular*}{\columnwidth}{@{\extracolsep{\fill}}lccl@{}}
    \hline
    Composition & E$^{\rm hull}$ & Decomposition products & Comments \\
    \hline
    Mg$_2$Mo$_3$O$_8$ & 51 & MoO$_2$ + MgO & Chemically synthesizable \\
    MgMo$_3$O$_8$ & 180 & MoO$_2$ + MgMoO$_4$ & -- \\
    Mo$_3$O$_8$ & 330 & MoO$_2$ + Mo$_8$O$_{23}$ & Naturally amorphous\cite{Vlek1977,Staples1951} \\
    \hline
  \end{tabular*}
\end{table*}

\begin{figure}[h!]
\centering
  \includegraphics[width=\columnwidth]{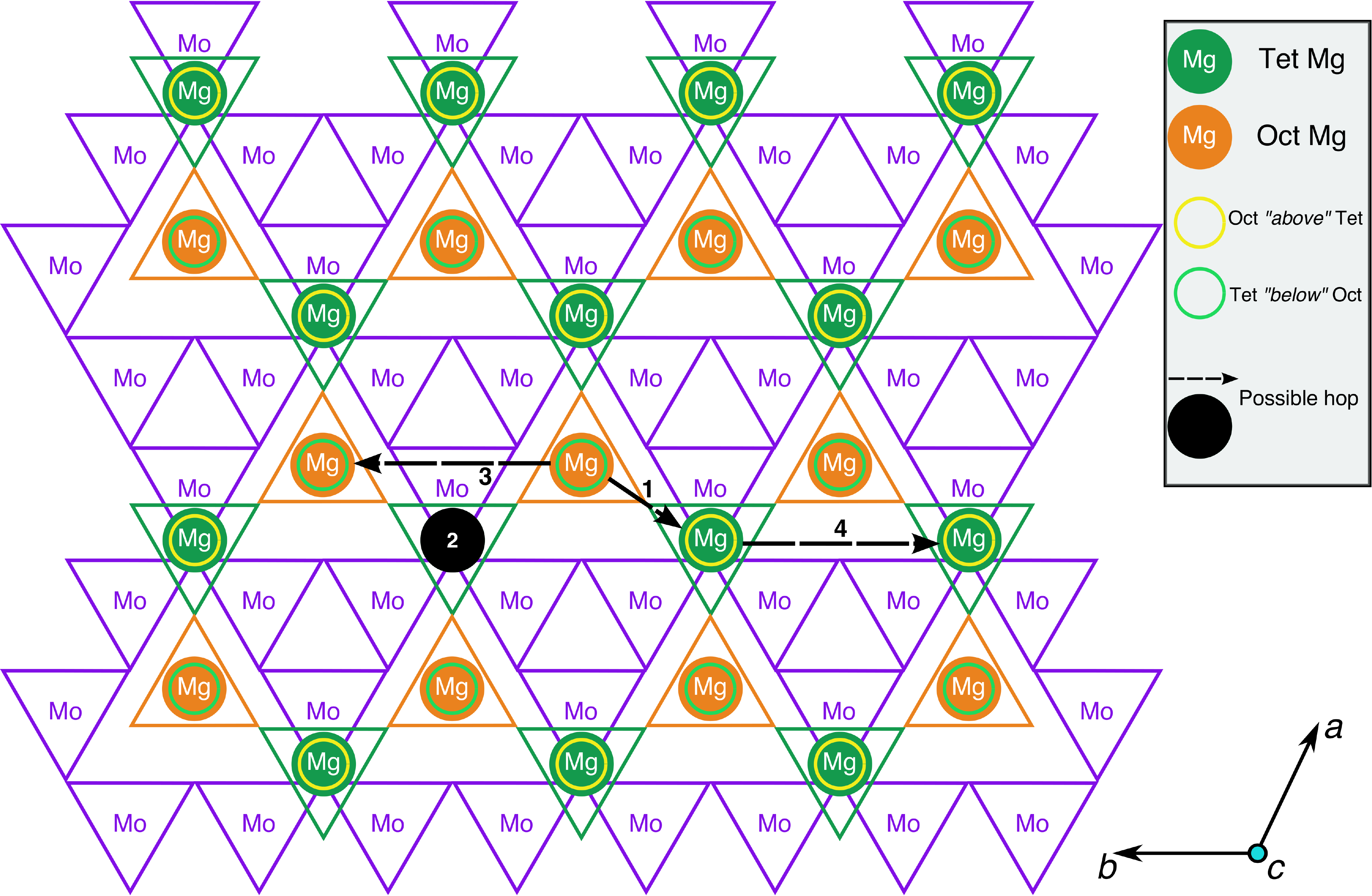}
  \caption{A 2D-view of the Mg$_2$Mo$_3$O$_8$ structure perpendicular to the layer spacing direction ($c$-axis) is shown. Purple, green and orange triangles indicate MoO$_6$ octahedra, Mg tetrahedra and Mg octahedra, respectively. The yellow and green circles correspond to octahedral and tetrahedral Mg atoms across a Mo-plane. The black circles and arrows indicate possible Mg$\rightarrow$Mg hops within the structure.}
  \label{fig:3}
\end{figure}

To evaluate Mg mobility in the Mg$_2$Mo$_3$O$_8$ structure, the possible Mg diffusion hops within the structure were determined. Being a layered structure, Mg$_2$Mo$_3$O$_8$ can be visualized on a 2D-plane, as shown in Figure~\ref{fig:3}, with octahedral Mo, tetrahedral Mg and octahedral Mg indicated by purple, green and orange triangles, respectively. The four possible Mg$\rightarrow$Mg hops that can occur in the structure are illustrated by the black circle and arrows in Figure~\ref{fig:3}. Three hops (black arrows) occur in the same Mg-plane and the fourth hop (black circle) moves Mg across a Mo-plane. The shortest hops (type 1 and 2) span $\sim$~3.38~{\AA} and $\sim$~4.33~{\AA}, respectively, and involve Mg migration from a tetrahedral site to an octahedral site (or vice-versa), while hops 3 and 4 are $\sim$~5.76~{\AA} in distance and involve Mg jumps between similarly coordinated sites (oct $\rightarrow$ oct or tet $\rightarrow$ tet). Although hops 3 and 4 are direct between octahedral or tetrahedral Mg sites, they are likely to be constituted by two consecutive hops of type 1 (i.e.\ an oct $\rightarrow$ tet hop followed by a tet $\rightarrow$ oct hop and vice-versa). Alternate routes for hops 3 and 4 are not possible due to intermediate Mg tetrahedral sites, which will face-share with MoO$_6$ octahedra and experience strong electrostatic repulsions as a result. Hence, hops 1 and 2 are the relevant Mg migration pathways that need to be considered in calculations.

\begin{figure}[h!]
\centering
  \includegraphics[width=\columnwidth]{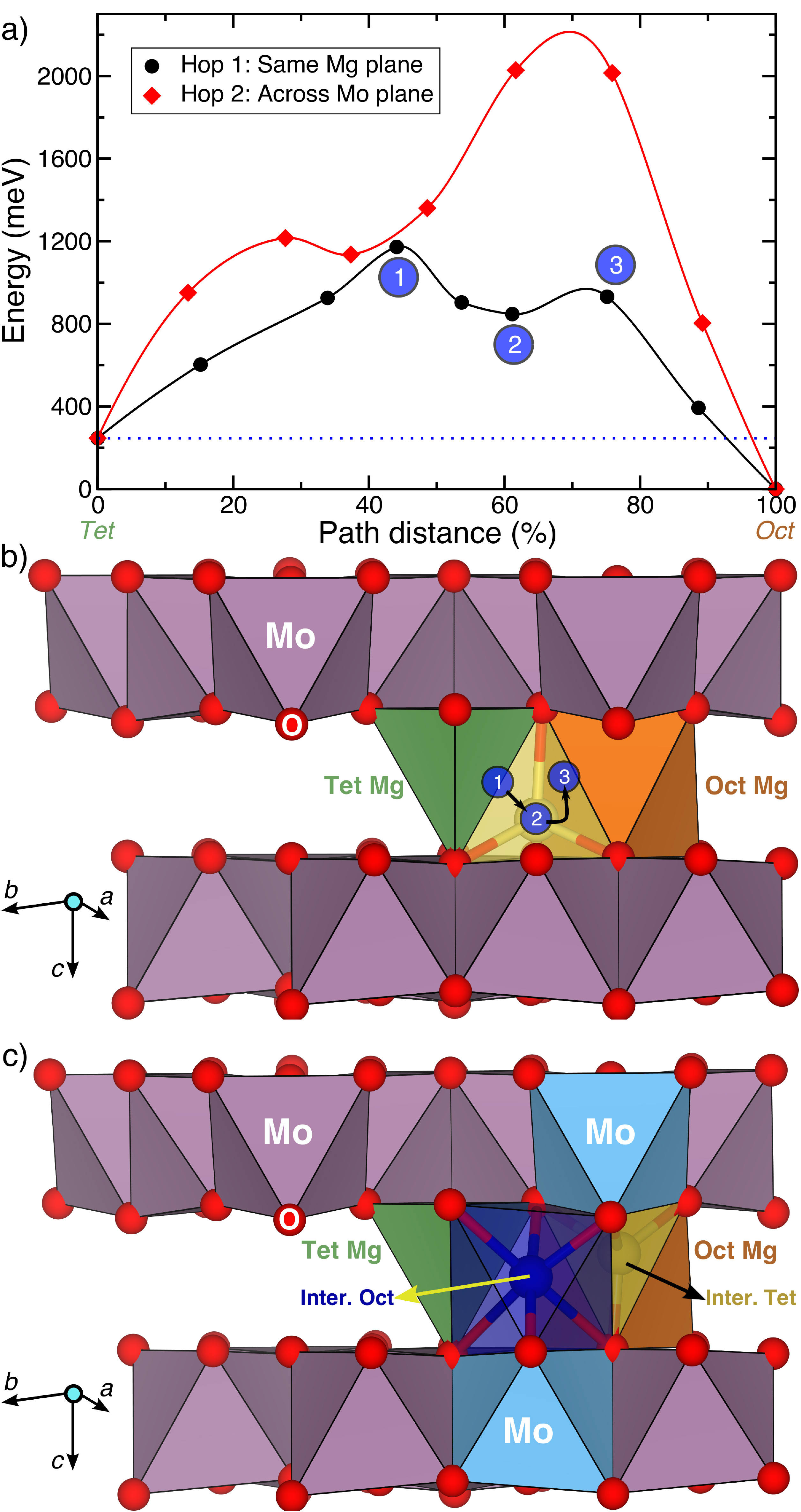}
  \caption{(a) The activation barrier for Mg diffusion along hops 1 and 2 in the Mg$_2$Mo$_3$O$_8$ structure, with the normalized path distance on the $x$-axis. (b) A closer view of hop 1, where the numbered circles correspond to various intermediate sites along the hop as labeled in (a). The intermediate tetrahedral site, which is edge-sharing with the stable tetrahedral site (green), is indicated in yellow. (c) An alternate pathway for hop 1 that involves intermediate octahedral (dark blue) and tetrahedral (yellow) sites, which are face-sharing with the stable tetrahedral (green) and octahedral (orange) sites, respectively. The intermediate sites in (c) also share a face with the MoO$_6$ octahedra (blue).}
  \label{fig:4}
\end{figure}

Figure~\ref{fig:4}a displays the calculated Mg migration barriers (at x$_{\rm Mg} \sim$~2 with dilute vacancy limit) along the hop 1 (black) and 2 (red) pathways, with the respective hop distances normalized on the $x$-axis. Both hops begin at a tetrahedral Mg and terminate at an octahedral Mg, explaining the difference in energy between the end points ($\sim$~250~meV). Notably, Mg mobility along both hops 1 and 2 is expected to be poor, given the large migration barriers ($\sim$~1200~meV and $\sim$~2000~meV for hops 1 and 2, respectively), compared to the 525~--~650 meV required for bulk Mg mobility at reasonable rates.\cite{Rong2015} The high migration barriers also explain the lack of electrochemical activity observed. Nevertheless, if any Mg migration is observed in the structure, the Mg$^{2+}$ ions are likely to diffuse along the in-plane hop 1 pathway.

While the high barrier for hop 2 is due to the strong electrostatic repulsion Mg experiences from Mo atoms as it passes through a triangular face of oxygen atoms across the Mo$_3$O$_8$ layer, a closer look into hop 1 is required to understand the large barriers. Visualization of the Mg migration along hop 1 is given in Figure~\ref{fig:4}b, with intermediate sites and their respective energies (in Figure~\ref{fig:4}a) indicated by the numbered circles. Sites 1, 2 and 3 respectively correspond to the O--Mg--O ``dumbbell" configuration,\cite{Rong2015} the intermediate metastable tetrahedral site (yellow) and the triangular face between the intermediate tetrahedral and stable octahedral sites. While site 3 (triangular face, Figure~\ref{fig:4}b) has an energy of $\sim$~685~meV with respect to the tetrahedral site (similar to $\sim$~600~--~800~meV observed in oxide spinels\cite{Liu2015}), the magnitude of the barrier is determined by site 1, where Mg is situated along an O--O bond (edge of the stable tetrahedron), in a dumbbell configuration. Previous evaluations of Mg migration through an O--O dumbbell hop for layered NiO$_2$ have reported high barriers ($\sim$~1400~meV),\cite{Rong2015} similar to the value reported in this work.
  

Although the O--Mg--O dumbbell hops are precluded from occurring in usual cathode materials\cite{Rong2015,SaiGautam2015} due to the presence of alternate low-energy pathways, no such possibility exists for Mg migration in the Mg$_2$Mo$_3$O$_8$ structure. For example, an alternate pathway for hop 1 that avoids the O--Mg--O dumbbell is shown in Figure~\ref{fig:4}c. The intermediate octahedral (dark blue) and tetrahedral (yellow) sites in Figure~\ref{fig:4}c share a triangular face with the stable tetrahedral (green) and octahedral (orange) sites, respectively. Additionally, each intermediate site also shares a triangular face with a MoO$_6$ octahedron (blue polyhedron, Figure~\ref{fig:4}c). While an intermediate Mg octahedron that face-shares with a higher valent transition metal octahedron need not preclude Mg migration, the intermediate tetrahedral site (yellow site, Figure~\ref{fig:4}c) will experience much stronger electrostatic repulsion from the face-sharing MoO$_6$ octahedron, subsequently increasing its energy and preventing any potential Mg migration. Indeed, Mg migration calculations initializing hop 1 as displayed in Figure~\ref{fig:4}c relax to a path similar to the O--Mg--O hop (Figure~\ref{fig:4}b) with a similar barrier ($\sim$~1150~meV, Figure~S5). Notably, scenarios involving a Mg$^{2+}$ ion diffusing through an intermediate (tetrahedral) site that face-shares with a transition metal polyhedron lead to high migration barriers in oxides (e.g., high Mg barriers in layered NiO$_2$\cite{Rong2015}), while analogous trends have been observed for Li-diffusion in disordered rock-salt structures.\cite{Urban2014}  Thus, the high Mg migration barrier in Mg$_2$Mo$_3$O$_8$ can be attributed to the intermediate O--Mg--O dumbbell configuration, which occurs in the absence of alternate low energy pathways. This indicates the importance of intermediate sites along a diffusion path, determined by the specific topology of cation sites in an anion lattice, in addition to the occurrence of the mobile cation with a non-preferred coordination and a preferentially coordinated metastable site.\cite{Rong2015}


One of the challenges towards the development of high energy density secondary Mg batteries is the design of an ideal positive electrode, which can reversibly intercalate Mg at a high voltage with high capacities at reasonable rates. The Mg$_2$Mo$_3$O$_8$ structure used in this study was primarily motivated by the presence of Mo$_3$ clusters (similar to the Mo$_6$ clusters in the Chevrel-positive electrodes) and the occurrence of Mg in a non-preferred tetrahedral coordination (satisfying one of the design rules known in literature\cite{Rong2015}). While Mg could be chemically extracted from the structure, albeit with significant amorphization, no electrochemical activity was observed. Further analysis using first-principles calculations revealed high E$^{\rm hull}$ values (structural instability) at low Mg content and high Mg migration barriers (poor bulk Mg mobility in the structure), explaining the aforementioned experimental observations. The high activation barrier for Mg diffusion in Mg$_2$Mo$_3$O$_8$ arises from the O--Mg--O dumbbell hop, reflecting the impact of intermediate sites along a diffusion pathway besides cation coordination preferences. Thus, in searches of high Mg-mobility oxide positive electrodes, a careful analysis of the diffusion pathway and the topology of cation sites is advantageous - such as identifying low-energy intermediate sites - in addition to the requirement of Mg being found in a non-preferred coordination environment.\cite{Rong2015} Such understanding of Mg diffusion pathways will help to find suitable positive electrodes for multivalent batteries.

\begin{acknowledgements}
This work was fully supported by the Joint Center for Energy Storage Research (JCESR), an Energy Innovation Hub funded by the U.S. Department of Energy, Office of Science and Basic Energy Sciences. The authors would like to thank the National Energy Research Scientific Computing Center (NERSC) for providing computing resources. Natural Sciences and Engineering Research Council of Canada (NSERC) is acknowledged by LFN for a Canada Research Chair.
\end{acknowledgements}

\bibliography{library}

\end{document}